# Robust Causal Transform Coding for LQG Systems with Delay Loss in Communications

Mohammad Kazem Izadinasab[*], Amir Homayoun Bahari Sani[**], Farshad Lahouti[+], and Babak Hassibi[+]

[*] Department of Electrical and Computer Engineering, University of Waterloo
[**] School of Electrical and Computer Engineering, College of Engineering, University of Tehran
[+] Department of Electrical Engineering, California Institute of Technology
mkizadinasab@uwaterloo.ca, bahari@ut.ac.ir, [lahouti, hassibi]@caltech.edu

*Abstract*— **A networked controlled system (NCS) in which the plant communicates to the controller over a channel with random delay loss is considered. The channel model is motivated by recent development of tree codes for NCS, which effectively translates an erasure channel to one with random delay. A causal transform coding scheme is presented which exploits the plant state memory for efficient communications (compression) and provides robustness to channel delay loss. In this setting, we analyze the performance of linear quadratic Gaussian (LQG) closed-loop systems and the design of the optimal controller. The design of the transform code for LQG systems is posed as a channel optimized source coding problem of minimizing a weighted mean squared error over the channel. The solution is characterized in two steps of obtaining the optimized causal encoding and decoding transforms and rate allocation across a set of transform coding quantizers. Numerical and simulation results for Gauss-Markov sources and an LQG system demonstrate the effectiveness of the proposed schemes.**

## I. INTRODUCTION

It is now widely recognized that many of the future applications of systems and control, such as those that arise in the context of cyber-physical systems, will pertain to problems of distributed estimation and control of multiple agents such as sensors and actuators over unreliable networks. While substantial research has been done on ways of constructing channel codes to facilitate reliable communications in network control systems (NCS), the design of source codes for rate efficient communications in such systems is much less explored. This paper presents the design of a transform-based source coding scheme for networked control systems communicating over channels with delay loss.

In NCS, one needs to deal with both the typical real-time constraints in control systems and the underlying unreliability of communications in a simultaneous and systematic way. A general framework to study such problems is introduced in [5] in the context of interactive communications over noisy channels. To solve the problem a new coding paradigm called "tree coding" is presented [5]-[7]. Subsequently in [8] the problem of stabilizing a plant over noisy links using tree codes is considered. To this end, they recognized that to stabilize an unstable plant over a noisy channel one needs real-time encoding and decoding, and a reliability that increases exponentially with delay. In [9], the existence with "high probability" of "linear" tree codes is proven and explicit codes with efficient encoding and decoding for the erasure channel are presented. The tree codes effectively replace a lossy link with a (asymptotically lossless) link with random delay.

In [4], a linear quadratic Gaussian (LQG) control problem where there is a lossy channel between the plant and the controller is considered. Under some rather general conditions, they show that the optimal LQG controller "separates" into a state-feedback controller and a "causal source coder" where the objective is to causally minimize a weighted mean-square-error of the plant state (the particular weighting coming from the LQG cost and plant dynamics). While this is quite nice, little is known about constructing appropriate causal source coding schemes.

Transform source coders ideally whiten the source signals and produces uncorrelated transform coefficients, which are more appropriate for quantization. Such a transform, which is known as the Karhunen Loeve transform (KLT), has the eigenvectors of the autocorrelation matrix of the source as its bases. Taking advantage of transform codes in controls applications calls for their equivalent causal counterparts. An interesting point is that one may still design a causal transform that optimally de-correlates the signal and hence achieves a coding gain similar to what is expected of KLT. The penalty though is that the transform will no longer be orthogonal and we require a more delicate quantization architecture[10]-[12].

In this paper, a causal transform code with vector quantization is presented for LQG closed-loop systems where the plant state is communicated over a channel with random delay. We analyze the system performance and the design of the optimal controller in this setting. The code is designed to minimize a weighted squared error measure and hence amounts to a joint source and channel coding or a channel optimized source coding scheme. To the best knowledge of the authors, this is the first work in this direction.

This paper is organized as follows. Section II presents the preliminaries including the system and channel models. Section III presents the structure of the proposed robust causal transform coding schemes. Section IV is devoted to the optimized design of controller and robust source coding. Performance evaluations of the proposed schemes in isolation and in the context of an LQG system are presented in Section V. Section VI concludes the paper.

## II. PRELIMINARIES

Figure 1 depicts the system model under consideration. We consider the following linear quadratic Gaussian plant:

$$x_{t+1} = Fx_t + Gu_t + w_t \quad (1)$$
$$y_t = Cx_t + v_t$$

where $x_t \in \mathbb{R}^d$ is a vector of states, $u_t \in \mathbb{R}^n$ is a vector of control signal, and $y_t \in \mathbb{R}^m$ is a vector of plant output. The terms $w_t \in \mathbb{R}^d$ and $v_t \in \mathbb{R}^m$ are independent and identically distributed (iid) zero-mean Gaussian vectors modeling the plant and measurement noise, respectively, with covariance matrices $K_w$ and $K_v$. We have $F \in \mathbb{R}^{d \times d}$, $G \in \mathbb{R}^{d \times n}$, $C \in \mathbb{R}^{m \times d}$ and the pair of $(F, G)$ is controllable.

Assuming the system state is to be communicated over the channel, the encoder maps each continuous plant state vector to an index in a finite alphabet. More specifically, the plant state is sampled with period $T_s$ and is encoded with a causal transform code $A$ of length $N$ and quantizer set $Q$. We have

$$x_c = \mathcal{E}_{A,Q}(x) \quad (2)$$

where $x_c$ is the encoder output that represents a reconstruction codevector, whose index is communicated over a lossy channel. In Section IV, we elaborate on the optimized design of the causal transform code.

We consider a channel model with random independent delay. Each encoder output index when communicated over the channel is subject to a time varying random independent delay $\delta$ which follows an exponential distribution with mean $1/\lambda$,

$$f_\delta(\delta) = \lambda e^{-\lambda \delta}. \quad (3)$$

Other models of channels with delay are available in [13].

The decoder maps back the indices received in time over the channel to a signal in $\mathbb{R}^m$, i.e., $\mathcal{D}: \mathcal{I} \cup \{\text{Erasure}\} \to \mathbb{R}^m$. In parallel to encoding, the decoding involves a reverse quantization that uses the reproduction codebook $\mathcal{C}$ and a transformation $\hat{A}$ which is not necessarily equal to $A$ because of the channel. Hence, we have

$$\hat{x} = \mathcal{D}_{\hat{A},\mathcal{C}}(\tilde{x}_c) \quad (4)$$

where as shown in Figure 1, $\tilde{x}_c$ is the channel output.

If a decoding delay of $\Delta$ is acceptable, any communicated index that is not available at the time of signal reconstruction is assumed erased. Since decoding of each sample relies on (up to) the past $N$ indexes, while a given index may be determined erased at a given decoding time, it may arrive for the subsequent decoding. The channel state at time $i$ is then described by a binary $N$ dimensional vector $[b_{ij}]_{j=1}^N$, which indicates the availability or unavailability of each of the indices at the receiver with a one or zero, respectively.

When there is no channel between plant and controller, i.e., under full state observation, the optimal control policy is the certainty equivalent control. The addition of communication channel to the system changes the problem to partially observed problem [4]. The LQG cost when plant states are sent over the channel turns into the following performance metric

$$\lim_{K \to \infty} \sup \frac{1}{K} \sum_{t=0}^{K-1} E_{X,B}[\hat{x}_t' R \hat{x}_t + u_t' S u_t] + E_{X,B}[e_t' R e_t] \quad (5)$$

where $R$ and $S$ are positive definite matrices with appropriate dimensions and the pair of $(F, R^{\frac{1}{2}})$ is observable [4]. Moreover, $\hat{x}_t$ is the decoder output, and the vector $e_t = x_t - \hat{x}_t$ indicates the state estimation error. Note that if the plant output is sent over the channel instead of the states, $R$ is replaced with $\bar{R} = C'RC$ in (5), where $C$ describes how states and plant outputs are related in (1).

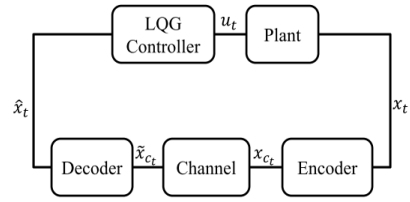

Fig. 1. System Model

## III. CAUSAL TRANSFORM CODING STRUCTURE

In this Section, we elaborate on the structure of causal transform code in an LQG networked control system. We consider a causal transform coding scheme of length $N$. The efficiency in such settings is characterized by the trade-off of rate and distortion that is typically quantified by mean squared error.

Figure 2 depicts the structure of the causal transform code. Within this structure, the encoder and decoder are designed by identifying the encoding matrix $A$, the decoding matrix $\hat{A}$, and $N$ vector quantizers and their rates $r_i, i = 1, \ldots, N$. The encoding matrix $A$ is an $mN \times mN$ unit diagonal lower triangular coding matrix

$$A = \begin{pmatrix} I_{m \times m} & 0 & \cdots & 0 & 0 \\ A_{21} & I_{m \times m} & & 0 & 0 \\ \vdots & & \ddots & & \vdots \\ A_{(N-1)1} & A_{(N-1)2} & & I_{m \times m} & 0 \\ A_{N1} & A_{N2} & \cdots & A_{N(N-1)} & I_{m \times m} \end{pmatrix},$$

where

$$A_{ji} = \begin{pmatrix} \alpha_{ji,1} & 0 & \cdots & 0 \\ 0 & \alpha_{ji,2} & \cdots & 0 \\ \vdots & 0 & \ddots & \vdots \\ 0 & 0 & \cdots & \alpha_{ji,m} \end{pmatrix}.$$

$I_{m \times m}$ is an $m \times m$ identity matrix. The decoding matrix $\hat{A}$ is another $mN \times mN$ unit diagonal lower triangular matrix with the same structure as $A$ where:

$$\hat{A}_{ji} = \begin{pmatrix} \alpha'_{ji,1} & 0 & \cdots & 0 \\ 0 & \alpha'_{ji,2} & \cdots & 0 \\ \vdots & 0 & \ddots & \vdots \\ 0 & 0 & \cdots & \alpha'_{ji,m} \end{pmatrix}$$

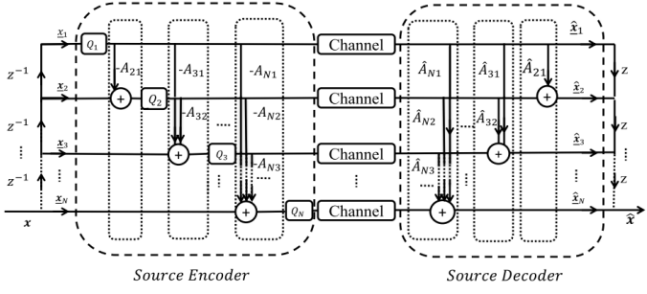

Fig. 2. Structure of encoder, channel and decoder.

The quantization noise may be considered as an additive noise vector $q$ of size $mN \times 1$. The encoding operation in Figure 2 may be described by

$$x_c = A^{-1}(x + q), \qquad (6)$$

where $x$ and $x_c$ are the $mN$-dimensional vectors of successive encoder input and output symbols, respectively.
Each vector quantizer is allocated $r_i$ bits satisfying

$$\frac{1}{N}\sum_{i=1}^{N} r_i = r, \qquad (7)$$

where $r$ is the average encoder bitrate.
At the decoder, the maximum allowable delay tolerated for reconstruction of each component is defined based on the requirements of the control system. Because of the ladder structure, the construction and reconstruction of each component affects the subsequent ones. At the time that the $i$'th element of the transform coding frame is to be reconstructed, the $j$'th element ($j \leq i$) has a maximum allowable delay of $\Delta + (i-j)T_s$, and as a result $pr\{b_{ij} = 0\} = pr\{\delta_{ji} > \Delta + (i-j)T_s\}$. For our channel model, based on (3) we have:

$$Pr\{\delta_{ji} > \Delta + (i-j)T_s\} = e^{-\lambda(\Delta+(i-j)T_s)}. \qquad (8)$$

We define $pr\{b_{ii} = 0\} = pr\{\delta_{ii} > \Delta\} = p$ and refer to it as delay violation probability in simulations. Note that for the reconstruction of elements that appear later in the transform coding frame, greater delay in reception of earlier elements could be tolerated.

To describe the decoder output, we present the channel state information (discussed in Section II) within a transform coding frame with the following matrix:

$$B = \begin{pmatrix} B_{11} & 0 & \cdots & 0 & 0 \\ B_{21} & B_{22} & & 0 & 0 \\ \vdots & & \ddots & & \vdots \\ B_{(N-1)1} & B_{(N-1)2} & \cdots & B_{(N-1)(N-1)} & 0 \\ B_{N1} & B_{N2} & \cdots & B_{N(N-1)} & B_{NN} \end{pmatrix},$$

where

$$B_{ij} = \begin{pmatrix} b_{ij} & 0 & \cdots & 0 \\ 0 & b_{ij} & \cdots & 0 \\ \vdots & 0 & \ddots & \vdots \\ 0 & 0 & \cdots & b_{ij} \end{pmatrix} \qquad (9)$$

where as stated $b_{ij} \in \{0,1\}, 1 \leq j \leq i \leq N$ is a Bernoulli random variable indicating the (un)availability of transform coding element $j$ at the decoder for the reconstruction of element $i$; $b_{ij} = 0, 1 \leq i < j \leq N$.

It is straightforward to verify that the $mN$-dimensional reconstructed vector at the receiver may now be described as follows

$$\hat{x} = (\hat{A} \circ B)x_c = Hx_c = HA^{-1}(x+q) \qquad (10)$$

where $H = \hat{A} \circ B$ and operator $\circ$ is element-wise Hadamard production.

## IV. CONTROLLER AND ROBUST SOURCE CODING DESIGNS

In this section, the design of controller and robust causal transform coding for closed-loop LQG problem is presented.

### A. Optimal Control Policy

Before dealing with the design, we consider the following definition.

*Definition:* The control is said to have *no-dual effect* if for all $\{u_t\}$ and $\forall t$, we have

$$E\{e'_t e_t | \tilde{x}_c^t, u^{t-1}\} = E\{e'_t e_t | \tilde{x}_c^t\} \qquad (11)$$

which implies that control has no role in reducing state uncertainty at the decoder or equivalently control has no effect on the second term of (5). In (11), $u^{t-1} = (u_1, u_2, \ldots, u_{t-1})$. The following proposition elaborates on optimal control policy when the robust causal transform code is utilized over the delay loss channel model presented.

*Proposition 1:* For a closed-loop LQG system with causal transform coding accompanying fine quantization and decoder reconstruction in (10), the subsystem from encoder to the decoder could be modeled by a block fading channel with additive independent noise and the optimal control policy is certainty equivalent control law.

*Proof:* Considering equation (9), it is evident that the block consisting of the encoder, the channel and the decoder is a fading channel with additive noise

$$\hat{x} = HA^{-1}x + HA^{-1}q = H_{eq}x + n_{eq} \qquad (12)$$

where $H_{eq} = (\hat{A} \circ B)A^{-1}$ is a lower triangular fading matrix which is constant over a transform coding frame, $N$, but varies frame to frame. When fine quantization assumption is invoked, the quantization noise, $q$, is independent of the input signal, $x$. Hence, the RHS of (12) indicates a block fading equivalent channel model for the encoder, lossy channel and the decoder. Such a model satisfies the requirements of lemma 5.2.1 in [16], and as a result the no-dual effect holds. And hence, the optimal control policy is certainty equivalent control law $u_t = L\hat{x}_t$ [4] where,

$$L = -(G'PG + S)^{-1}G'PF \qquad (13)$$

in which $P$ is a matrix that satisfies the following Riccati equation

$$P = F'(P - PG(G'PG + S)^{-1}G'P)F + R. \quad (14)$$

∎

*Corollary 1:* For small delay violation probability, $p \to 0$, the robust causal transform encoder, the lossy channel and the corresponding decoder can be modeled as an equivalent channel with additive independent noise, i.e.,

$$\hat{x} = x + q.$$

*Proof:* In Proposition 1, as $p \to 0$, $B_{ij} \approx I, 1 \le j \le i \le N$ in (9) and $\hat{A} \approx A$ and $H_{eq} \approx I$.

∎

### B. Code Design Problem Formulation and Analysis

We here investigate the design criteria for the robust causal transform code based on the control policy prescribed in Proposition 1. We also analyze the performance of the closed-loop LQG system with robust causal transform coding.
We consider the partially observed LQG problem of (1). As stated in [4], this could be converted to a fully observed LQ problem with $\bar{w} = E[Fe_t + w_t | u, \tilde{x}_c^{t+1}]$. Hence, the covariance of this equivalent matrix is

$$K_{\bar{w}} = F'\Lambda F + K_w - \Lambda \quad (15)$$

where $\Lambda = \Lambda_t = \text{Cov}(e_t)$ and the optimal LQ cost is equal to $tr(PK_{\bar{w}})$.
Considering (5) and (15) we have

$$\lim_{K \to \infty} \sup \frac{1}{K} \sum_{t=0}^{K-1} E_{X,B}[\hat{x}_t'R\hat{x}_t + u_t'Su_t] \quad (16.a)$$
$$+ E_{X,B}[e_t'Re_t]$$
$$= tr(PK_{\bar{w}}) + tr(R\Lambda) \quad (16.b)$$
$$= tr(P(F'\Lambda F + K_w - \Lambda)) + tr(R\Lambda) \quad (16.c)$$
$$= tr(PK_w) + tr((F'PF - P + R)\Lambda). \quad (16.d)$$

In fact, the effect of coding scheme on both terms of (16.a) has been collected in the second term of (16.d) to provide us with a criterion for code design. Based on this analysis, we consider the arithmetic mean of the weighted mean square error between the plant states and the decoder outputs as the joint source channel coding design criteria

$$\text{AM} - \text{WMSE} = \frac{1}{mN} E_{X,B}[\|x - \hat{x}\|_M^2] \quad (17)$$
$$= \frac{1}{mN} E_{X,B}[(x - \hat{x})'M(x - \hat{x})]$$

Comparing equations (16.d) and (17), according to causal transform coding structure the weight matrix is $M$ is given by

$$M = \begin{pmatrix} R_{eq} & 0 & \cdots & 0 \\ 0 & R_{eq} & \cdots & 0 \\ \vdots & 0 & \ddots & \vdots \\ 0 & 0 & \cdots & R_{eq} \end{pmatrix}$$

Where

$$R_{eq} = F'PF - P + R. \quad (18)$$

The weight matrix $M$ depends on the LQG cost and system matrices and indicates that while different components of a transform coding frame are equally weighted (over time), the elements within each vector may be weighted differently.
Formally for a given channel and channel code the desired robust transform code design problem is formulated as a distortion-rate optimization problem as follows

$$\min_{A,\hat{A},\{r_1,r_2,\dots,r_N\}} \frac{1}{mN} E_{X,B}[\|x - \hat{x}\|_M^2] \quad (19)$$
$$s.t.: \frac{1}{N}\sum_{i=1}^{N} r_i = r$$

This means specifically identifying the optimized causal transform $A_{mN \times mN}$, the corresponding decoding transform $\hat{A}_{mN \times mN}$, and the rates of the associated $N$ quantizers $\{r_1, r_2, \dots, r_N\}$ over the lossy channel.
The analysis in (16) and the Proposition 1 enable the analysis of the LQG closed-loop system performance with robust causal transform coding and in presence of delay loss in communications. The following Proposition presents this result.

*Proposition 2:* For a closed-loop LQG system with causal transform coding accompanying fine quantization and decoder reconstruction in (10), the LQG cost when the plant state is sent over the channel is given by

$$tr(PK_w) + tr(E_B[(I - H_{eq})'R_{eq}(I - H_{eq})]K_x) \quad (20)$$
$$+ tr(E_B[H'_{eq}R_{eq}H_{eq}]K_q)$$

where, $K_x$ and $K_q$ are plant state and quantization error covariance matrices.

*Proof:* From Proposition 1, we have

$$e = (I - H_{eq})x - n_{eq}$$

And

$$E_{X,B}[e'R_{eq}e] = E_{X,B}\left[\left((I - H_{eq})x - n_{eq}\right)'R_{eq}\left((I - H_{eq})x - n_{eq}\right)\right]$$
$$= E_{X,B}\left[x'(I - H_{eq})'R_{eq}(I - H_{eq})x\right]$$
$$-2E_{X,B}\left[x'(I - H_{eq})'R_{eq}n_{eq}\right] + E_{X,B}\left[n'_{eq}R_{eq}n_{eq}\right]$$
$$= E_{X,B}\left[x'(I - H_{eq})'R_{eq}(I - H_{eq})x\right]$$
$$-2E_{X,B}\left[x'(I - H_{eq})'R_{eq}n_{eq}\right] + E_{X,B}\left[q'H'_{eq}R_{eq}H_{eq}q\right]$$

The second term of above with fine quantization assumption is zero, hence

$$tr(R_{eq}\Lambda) = \lim_{K\to\infty} \sup \frac{1}{K}\sum_{t=0}^{K-1} E_{X,B}[e_t'R_{eq}e]$$
$$= tr\left(E_B[(I - H_{eq})'R_{eq}(I - H_{eq})]K_x\right)$$
$$+ tr(E_B[H'_{eq}R_{eq}H_{eq}]K_q)$$

where, $K_x$ and $K_q$ are the plant state and quantization error covariance matrices. Hence, the optimal LQG cost will be

$$tr(PK_w) + tr\left(E_B[(I - H_{eq})'R_{eq}(I - H_{eq})]K_x\right)$$
$$+ tr(E_B[H'_{eq}R_{eq}H_{eq}]K_q)$$

∎

Note that using Proposition 2, given the transform coding matrices, the LQG system parameters and the covariance of the quantization noise, one may numerically analyze the system performance. The statistics of the quantization error with fine quantization (asymptotic regimes) have been studied in, e.g., [17].

*Remark:* In the setting of Corollary 1, when the delay violation probability is small, the LQG cost in Proposition 2 simplifies as follows

$$tr(PK_w) + tr(R_{eq}K_q)$$

### C. Code Design Scheme

The robust source coding design optimization problem in (19) does not lend itself to a direct closed-from solution, due to the nonlinearity of the quantizers and the closed-loop structure of the system. Instead, we take a two-step approximate solution. To obtain $A^*$ and $\hat{A}^*$, we consider fine quantization and invoke a numerical optimization approach in line with the non-gradient direct search method of Hooks and Jeeves [14]. Next, given the optimized transform matrices, we obtain the optimized rate allocation scheme to the set of quantizers.

A very good initial point for the search algorithm is the causal transform corresponding to the ideal (lossless) communication aka the Prediction-based lower triangular transform (PLT) scheme [12]. Once optimized solutions of $A^*$ and $\hat{A}^*$ are available in (19), it is straight forward to verify that the quantization rate allocation problem may be posed as a *weighted quantization error* [3] problem. We have

$$\min_{\{r_1,\dots,r_N\}} \frac{1}{mN} E_B\{q'Wq\}$$
$$s.t.: \frac{1}{N}\sum_{i=1}^{N} r_i = r \quad (21)$$

where, $W = (HA^{-1})'M(HA^{-1})$. For a symmetric and positive definite $W$, decomposing $W = Z'Z$ with $Z$ a lower triangular matrix, (21) coverts to an *unweighted quantization error* problem. The optimal bit allocation with fine quantization assumption using Karush–Kuhn–Tucker (KKT) conditions is

$$r_i^* = r + \frac{1}{2}\log_2 \frac{\hat{\sigma}_i^2}{(\prod_{k=1}^{N}\hat{\sigma}_k^2)^{\frac{1}{N}}}, i = 1,2,\dots,N \quad (22)$$

where, $\hat{\sigma}_i^2 = \left[\det\left(E_B\left[\text{Cov}\left(d_{i_{eq}}\right)\right]\right)\right]^{\frac{1}{m}}$, $d_{i_{eq}} = Zd_i$ and $d_i$ is the vector of inputs of the $i$-th quantizer. To derive (22), it has been assumed that the quantizer dimensions and input distributions are the same (here Gaussian). Note that in presence of the channel, even if we have $M = I$, we still face a weighted quantization error problem as $W \neq I$.

Note that in the lossless scalar case, the rate allocation in (22) coincides with what was obtained in [12] for the PLT structure. It is noteworthy that the optimized rate allocations in (22) are obtained without regard to the non-negativity or possible integer constraints on quantization rates. In practice, one may resort to heuristic techniques to impose such constraints on the obtained results. In [2], related algorithms for the case of classic transform codes (non-causal) are reported.

### D. Structured Encoding and Decoding Matrices

In addition to optimized matrices for Robust Causal Transform Code (RC-TC), structured coding and decoding matrices could also be considered for reduced design and implementation complexity. One interesting structure is the unit diagonal lower triangular and Toeplitz matrices for coding and decoding. Such an encoding matrix is given by

$$A_{Toeplitz} = \begin{pmatrix} I_{m\times m} & 0 & \dots & 0 & 0 \\ A_{21} & I_{m\times m} & \dots & 0 & 0 \\ A_{31} & A_{21} & \ddots & 0 & 0 \\ \vdots & A_{31} & A_{21} & I_{m\times m} & 0 \\ A_{N1} & \dots & A_{31} & A_{21} & I_{m\times m} \end{pmatrix}$$

We refer to this as the Robust Toeplitz Causal Transform Code (RTC-TC). In RC-TC, because of the unit diagonal lower triangular structure there are $m(N^2 - N)$ optimization variables whereas in the case of RTC-TC there are only $2m(N - 1)$ variables. This difference in complexity is considerable for large $N$.

## V. PERFORMANCE RESULTS

In this Section, we study the performance of the proposed robust causal transform code (RC-TC) and the robust Toeplitz causal transform code (RTC-TC) for communication of a Gauss-Markov source over a lossy channel. We also investigate the performance of the proposed schemes in an LQG system. For comparisons, we consider plain quantization (no transform coding), and PLT [12] schemes.

### A. Robust Causal Transform Coding

We experimented with two scalar ($m = 1$) GM sources with orders 1 and 10 [15]. We here report the first order case where the source is zero mean with unit variance and coefficient equal to 0.9. We compare the AM-MSE performance ($M = I$) for the four schemes. In the analysis, the covariance matrix $K_q$ is set based on $\sigma_{q_i}^2 = c2^{-2r_i}\sigma_{d_i}^2$ for a Gaussian source, where $c$ is a constant depending on the type of quantizer [1]. Figure 3 shows the analytic and simulated AM-MSE performance results for the source. One sees that the performance of PLT structure is better than that without coding. In addition, the proposed RC-TC and RTC-TC schemes still improve the performance. According to the conducted simulations, the transform coding schemes provide a higher gain for coding of sources with higher orders, which may be attributed to their stronger memory structure. It is noteworthy that the RTC-TC provides a performance very similar to that of the RC-TC, albeit with much smaller (design) complexity especially for large values of $N$. herefore, in subsequent simulations we will focus on the RTC-TC.

Our experiments (not reported here) reveal that RTC-TC and RC-TC with higher transform dimensions have better performance. A larger $N$ allows the transform to exploit the source dependencies more effectively and provides a greater opportunity for the delayed arrival of previous symbols at the cost of increased complexity.

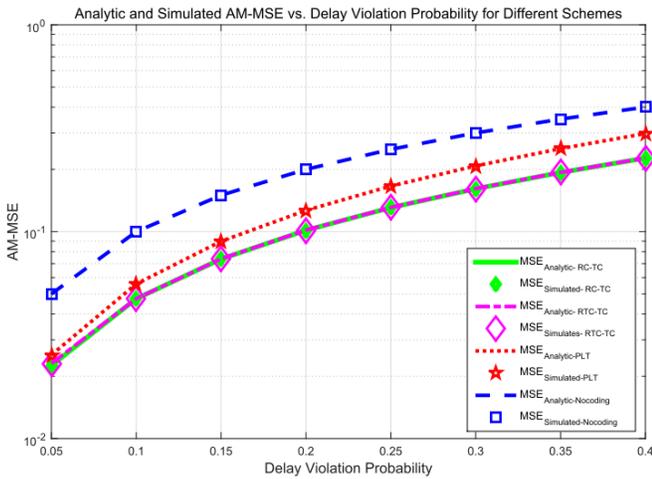

Fig. 3. Performance of different schemes for $N = 6, r = 5, \Delta = 50$ msec and $T_s = \frac{\Delta}{4}$.

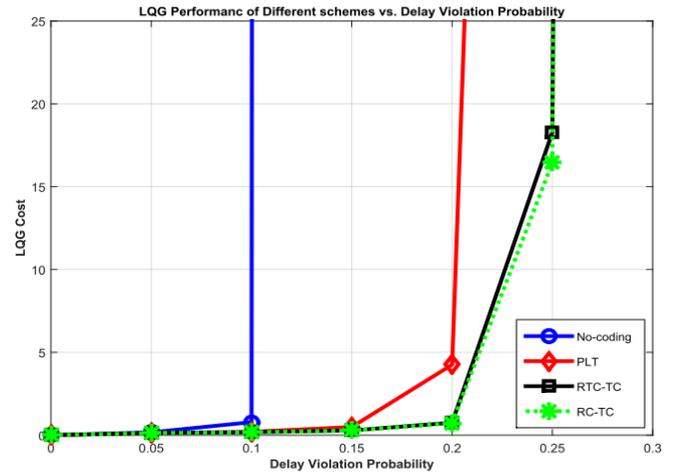

Fig. 4. LQG performance of different schemes as a function of delay violation probability.

### B. Robust Causal Transform Coding in NCS

In this part, we examine the performance of the proposed robust causal transform coding schemes in a closed-loop LQG system. We consider the following LQG system

$$\begin{cases} x_{n+1} = 1.49 x_n + 0.05 u_n + w_n \\ y_n = x_n + v_n \end{cases} \quad (23)$$

For this system, $R = 1$, $M = 1$ and variances of $w_n$ and $v_n$ are $0.01$ and $0.001$, respectively. For numerical optimization of causal transform coding matrices, the plant states may be modeled approximately by a first order GM source with coefficient $0.8677$.

Figure 4 depicts the LQG cost of the system in (23) for the four schemes and $N = 8$, $r = 5$, $\Delta = 0.05$sec and $T_s = \frac{\Delta}{4}$. As evident, the proposed schemes significantly reduce the LQG cost of the system and enhance the resilience of the system to the delay loss over the channel.

## VI. CONCLUSIONS

The main purpose of this paper has been to design robust causal transform codes for networked control systems communicating over a channel with delay loss. Two channel optimized coding scheme were proposed which provide a trade-off of performance and complexity of design and implementation. The suggested schemes exploit the dependencies in the plant state for efficient communications (compression) and provide robustness to channel loss. We also analyzed the performance and the design of optimal controllers with fine quantization. Numerical results and simulations demonstrate the effectiveness of the presented schemes for closed-loop LQG systems and the effect of design parameters.

### ACKNOWLEDGMENT

This work was supported in part by the National Science Foundation under grants CCF-1423663 and CCF-1409204, by a grant from Qualcomm Inc., by NASA's Jet Propulsion Laboratory through the President and Director's Fund, and by King Abdullah University of Science and Technology.

### REFERENCES

[1] N. S. Jayant and P. Noll, *Digital Coding of Waveforms*, Englewood Cliffs, NJ:Prentice-Hall, 1954.

[2] K. Sayood, *Introduction to Data Compression*. San Francisco, CA: Morgan Kaufmann, 2000.

[3] A. Gersho and R. M. Gray, *Vector quantization and signal compression*, Springer, 1992.

[4] S. Tatikonda , A. Sahai and S. M. Mitter, "Stochastic linear control over a communication channel," *IEEE Trans. Autom. Control*, vol. 49, no. 9, pp.1549 -1561, 2004.

[5] L. Schulman, "Coding for interactive communication," *IEEE Transactions on Information Theory,* Vol. 42, pp. 1745–1756, Dec. 1996.

[6] M. Braverman and A. Rao, "Towards coding for maximum errors in interactive communication," *in Proc. of the 43rd annual ACM Symposium on Theory of Computing,* 2011.

[7] R. Gelles and A. Sahai, "Potent tree codes and their applications: Coding for interactive communication, revisited," *CoRR*, vol. abs/1104.0739, 2011.

[8] A. Sahai and S. Mitter, "The necessity and sufficiency of anytime capacity for stabilization of a linear system over a noisy communication link–part i: Scalar systems," *IEEE Transactions on Information Theory,* 52:3369–3395, August 2006.

[9] R. T. Sukhavasi and B. Hassibi, "Error correcting codes for distributed control," http://arxiv.org/abs/1112.4236l, February 2012.

[10] F. Lahouti and A. K. Khandani, "Sequential vector decorrelation technique," Technical Report UW-ECE-2001-3, 2001.

[11] S. M. Phoong and Y. P. Lin, "Prediction-based lower triangular transform," *IEEE Trans. Signal Process.,* Vol. 48, No. 7, pp. 1947–1955, July 2000.

[12] C.-C. Weng, C.-Y. Chen, and P. P. Vaidyanathan, "Generalized triangular decomposition in transform coding," *IEEE Trans. Signal Process.,* Vol. 58, No. 2, February 2010.

[13] J. Nilsson, Real-time control systems with delays, 1998 :Dept. Automatic Control, Lund Inst. Technol.

[14] R. Hooke and T.A. Jeeves, "Direct search solution of numerical and statistical problems," *Journal of the Association for Computing Machinery (ACM)*, 8(2):212–229, 1961.

[15] F. Lahouti and A. K. Khandani, "Reconstruction of predictively encoded signals over noisy channels using a sequence MMSE decoder," *IEEE Trans. Commun.*, pp.1292 -1301, 2004.

[16] D. P. Bertsekas. Dynamic Programming and Optimal Control. Athena Scientific ,2nd ed. , 2000.

[17] D. H. Lee and D. L. Neuhoff, "Asymptotic distribution of the errors in scalar and vector quantizers", IEEE Trans. Inform. Theory, vol. 42, pp.446 -460 1996.